\documentclass[10pt,leqno]{amsart}

\usepackage{graphicx}
\usepackage[numbers,sort&compress]{natbib}
\usepackage{multirow}
\usepackage{amsmath,amssymb,amsfonts,amsthm,mathrsfs}
\usepackage{textcomp,booktabs,makecell}
\usepackage[normalem]{ulem}
\usepackage{xcolor}
\usepackage{indentfirst,csquotes}
\usepackage{titlesec}
\usepackage{hyperref}

\titleformat{\section}[display]{\normalfont\Large\bfseries\centering}{}{0pt}{}
\titlespacing*{\section}{0pt}{1.5ex}{1ex}
\hypersetup{colorlinks=true,linkcolor=black,citecolor=black,urlcolor=black}

\title{Jupiter's Polar Vortex Structure and Drift Reveal Its Atmospheric Dynamics}
\author{Jiaru Shi}
\address{Department of Earth, Atmospheric and Planetary Sciences, Massachusetts Institute of Technology, Cambridge, MA, USA}
\email{sjiaru@mit.edu}
\author{Wanying Kang}
\address{Department of Earth, Atmospheric and Planetary Sciences, Massachusetts Institute of Technology, Cambridge, MA, USA}
\date{\today}

\begin{document}
\begin{abstract}
Jupiter's polar atmosphere hosts a westward-drifting vortex crystal: a stable arrangement of coherent cyclones embedded in an anticyclonic background. Dry models sustain multiple vortices in two regimes, but neither simultaneously reproduces the crystal structure and the observed drift. Strong atmospheric stratification yields a cyclone-dominated crystal, but the vortices drift too fast and in the wrong direction. Weak stratification yields the observed slow westward drift, but the vortices form a disordered ``soup'' instead of a crystal. At intermediate stratification, dry models sustain only a single vortex. We show that a convective feedback motivated by stratiform instability narrows this gap across the full range of stratification, in a different way in each regime. Temperature perturbations associated with cyclones reduce convective inhibition and enhance convective forcing over cyclonic regions. For weak stratification, this concentrates the forcing in cyclones and produces a cyclone-dominated vortex crystal while preserving the slow drift. For intermediate and large stratification, the feedback wraps each cyclone in an anticyclonic shielding layer. This helps sustain multiple vortices for intermediate stratification and helps reverse the vortex drift direction for strong stratification. 
\end{abstract}

\maketitle

\section*{Introduction}

At Jupiter's poles, a mysterious vortex crystal forms. It consists of several similarly sized cyclones concentrated near each pole, arranged in a crystal-like pattern, and drifting slowly westward \cite{Mura-Adriani-Bracco-et-al-2021:oscillations,Adriani-Mura-Orton-et-al-2018:clusters,Adriani-etal-2020:Two-year,Mura-etal-2022:Five-year}. Such a vortex configuration has not been observed on any other planet in the Solar System \cite{Wilson-1997:Martian,Baines-Momary-Fletcher-et-al-2009:saturns,Luz-etal-2011:venus,Teanby-etal-2017:Titan}, and it contrasts with the expectation that cyclonic vortices migrate poleward and merge under the influence of the planetary vorticity gradient due to planetary curvature \cite{McWilliams-Flierl-1979:evolution,Sutyrin-1988:Motion,Sutyrin-Flierl-1994:intense,Korotaev-Fedotov-1994:dynamics,Scott-2011:Polar,ONeill-Emanuel-Flierl-2015:polar,ONeill-Emanuel-Flierl-2016:Weak}.

Drawing on earlier studies of $\beta$-plane turbulence and vortex dynamics, researchers have proposed two mechanisms for sustaining Jupiter's polar vortex crystal:
\begin{itemize}
\item \textbf{Mechanism 1:} Turbulent mixing homogenizes potential vorticity in the polar regions, shutting down the driving force for the poleward migration of cyclones \cite{Rhines-Young-1982:homogenization,Marcus-Lee-1998:Jets,Dritschel-McIntyre-2008:multiple,Siegelman-Young-Ingersoll-2022:polar,Chen-Ingersoll-Li-2024:jupiter}.
\item \textbf{Mechanism 2:} Weak interior stratification (characterized by a small Rossby deformation radius $L_d$) suppresses interactions among vortices and with planetary vorticity \cite{Flierl-1977:Gulf,McWilliams-Flierl-1979:evolution,Polvani-McWilliams-Spall-1994:coherent,Brueshaber-Sayanagi-Dowling-2019:dynamical}.
\end{itemize}
These mechanisms were initially studied using models of varying complexity \cite{Brueshaber-Sayanagi-Dowling-2019:dynamical,Li-Ingersoll-Klipfel-et-al-2020:modeling,Siegelman-Young-Ingersoll-2022:polar,Chen-Ingersoll-Li-2024:jupiter}. Shi and Kang \cite{Shi-Kang-2026} unified them within a 1.5-layer quasi-geostrophic (QG) framework and identified the parameter regimes in which each dominates. They found that coherent vortex patterns form as long as the damping is sufficiently weak (i.e., the damping scale $L_\mu$ is large), and that the vortex pattern is largely determined by the ratio between the deformation radius $L_d$ and the zonostrophic scale $L_\gamma$.

When $L_d\gg L_\gamma$, multiple vortices can be sustained by Mechanism~1. These vortices exhibit strong cyclone--anticyclone asymmetry and a regular crystal structure \cite{Siegelman-Young-Ingersoll-2022:polar,Chen-Ingersoll-Li-2024:jupiter,Shi-Kang-2026}, closely resembling the polar vortex clusters observed on Jupiter \cite{Adriani-Mura-Orton-et-al-2018:clusters}. However, because of strong advection by nearby vortices, such crystals preferentially drift eastward at very high speeds ($\sim$40 rotation periods per year) \cite{Siegelman-Young-Ingersoll-2022:polar}. In contrast, Jupiter's polar vortex crystal drifts slowly westward, at an average rate of $\sim 3$--$7^\circ$/yr \cite{Mura-Adriani-Bracco-et-al-2021:oscillations}. Indeed, Gavriel and Kaspi \cite{Gavriel-Kaspi-2025:dynamical} found that matching the observed drift speed requires a deformation radius of about 190~km over the north pole and 320~km over the south pole.

At such small $L_d$, stable vortex crystals cannot form \cite{Chen-Ingersoll-Li-2024:jupiter,Shi-Kang-2026}. Instead, when $L_d<L_\gamma$, Mechanism~2 prevents vortex merging by weakening vortex--vortex interactions. Because the vortices barely feel one another, cyclones and anticyclones of comparable strength \cite{Brueshaber-Sayanagi-Dowling-2019:dynamical,Shi-Kang-2026} arrange themselves randomly into a vortex ``soup''. This state drifts in the right direction but has the wrong appearance.

Getting beyond spontaneous pattern formation from random forcing, people find that prescribing an anticyclonic shielding layer around each cyclone can introduce a repellent force and thereby sustain a vortex crystal \citep{Li-Ingersoll-Klipfel-et-al-2020:modeling,Gavriel-Kaspi-2021:number,Gavriel-Kaspi-2023:westward}. Such an anticyclonic shielding layer can form as a vortex migrates from low latitude under $\beta$ drift \cite{Korotaev-Fedotov-1994:dynamics,Carruthers-etal-2023:shielding}. 
A three-dimensional model and a forced-dissipative shallow-water model have also produced vortex crystals dynamically \cite{Cai_2021,Cai-2024-examination,ruan2026jovianpolarvortexgeneration}. In both studies, however, the simulated crystals occupy the entire computational domain. It therefore remains an open question whether such well-organized structures can persist over a much larger portion of the planetary surface.

In this work, we build upon the dry QG framework \cite{Shi-Kang-2026} to test whether adding a moist-convective feedback can resolve the difficulty of simultaneously capturing vortex morphology and drift speed. In particular, we ask whether the feedback can (i) introduce cyclone-anticyclone asymmetry in the Mechanism~2-dominated regime, so that the vortex ``soup'' becomes more like a vortex crystal while keeping its drift direction and speed, and (ii) reverse the drift direction and slow the drift speed when applied to the Mechanism~1-dominated regime.

As we show later, after building this feedback into the QG framework of Shi \& Kang \cite{Shi-Kang-2026}, we reproduce crystal-like cyclone-dominated vortex patterns that drift slowly westward, as observed, in both the Mechanism~1-dominated and Mechanism~2-dominated regimes. For the Mechanism~2-dominated regime, we further derive quantitative relationships that predict the cyclone--anticyclone energy partition, vortex drift speed, inter-vortex spacing, and root-mean-square wind speed from the model parameters, and we validate them against numerical simulations. Because these relationships link observable quantities to model parameters, they allow us to constrain several internal properties of Jupiter that those parameters represent.


\begin{figure}[ht!]
\centering
\includegraphics[width=0.7\linewidth]{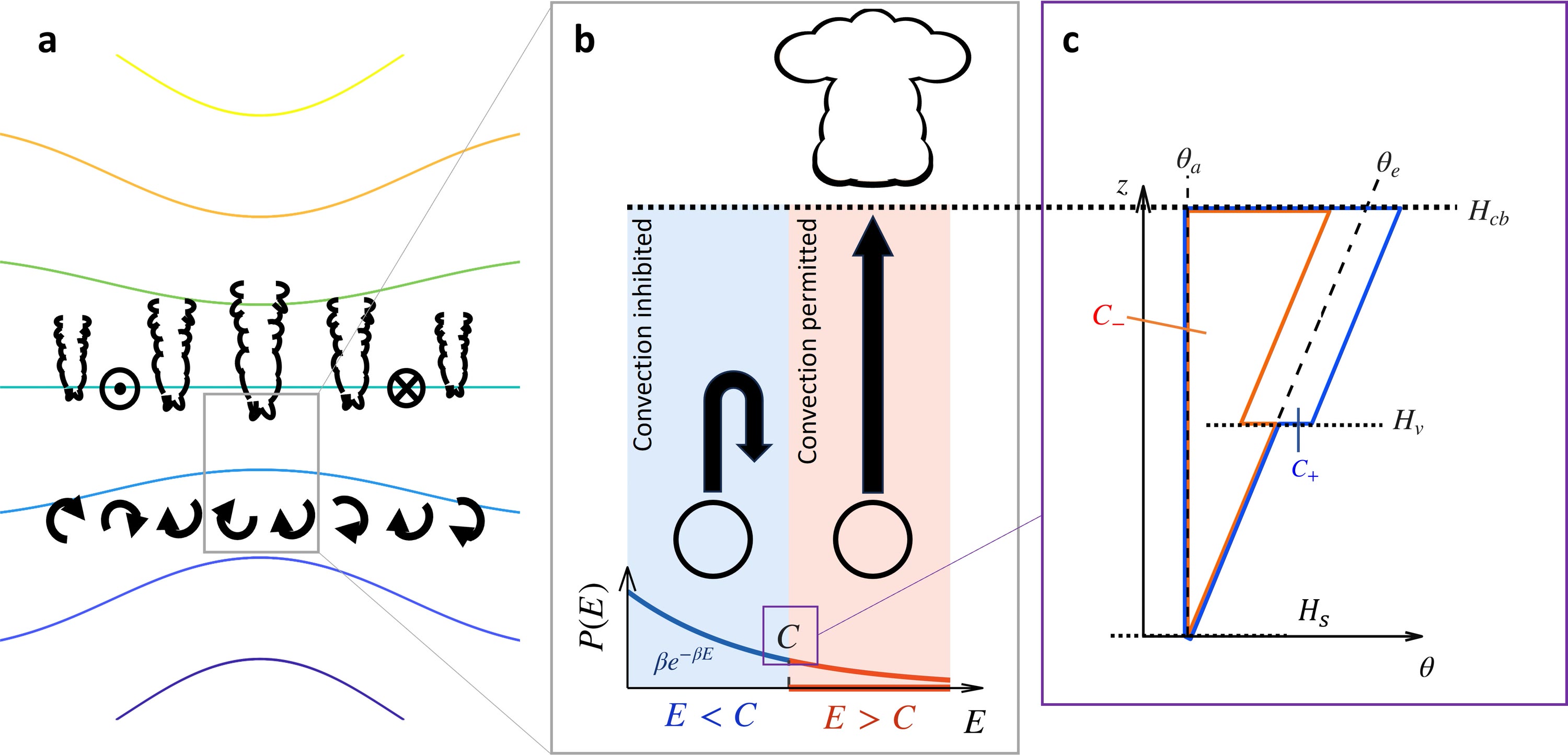}
\caption{A schematic of stratiform instability. In panel a, solid colored lines indicate isentropes perturbed by a vortex, black eddies represent turbulence in the sub-cloud layer, and clouds above denote convective forcing. In panel b, air parcels (black circles) rise with initial energy $E$, while only parcels whose initial energy is higher than the convective inhibition $C$ reach the cloud base at $H_{cb}$. In panel c, the convective inhibition is illustrated by the area of the triangle bounded by the air-parcel potential-temperature profile $\theta_a$, the environmental potential-temperature profile $\theta_e$, and the cloud-base depth $H_{cb}$. $H_s$ and $H_v$ are the source level of air parcels and the level at the base of the vortex, respectively. The presence of a vortex perturbs $C$ by changing the profile of $\theta_e$, producing either the smaller value $C_-$ or the larger value $C_+$. If the vortex is cyclonic, $C$ is reduced to $C_-$, making the convective forcing stronger.}
\label{fig:schematics}
\end{figure}

\section*{Methods}\label{sec:methods}

\subsection*{The stochastically forced QGPV equations}

The atmospheric motions on rapidly rotating gas giants, such as Jupiter and Saturn, are well described by a quasi-geostrophic framework. Using a stochastically forced QG model, Shi and Kang \cite{Shi-Kang-2026} identified four possible vortex morphologies and the parameter regimes in which each can emerge. Their governing equation is the QG potential vorticity (PV) equation on a $\gamma$-plane:
\begin{equation}
\label{eq:PVeq}
D_t\Big(q - \frac{1}{2}\gamma r^2\Big) = F\eta - \mu q, \quad
q = \nabla^2\psi - L_d^{-2}\psi,
\end{equation}
where $q$ denotes the potential vorticity, $\psi$ is the streamfunction, and $\nabla^2\psi$ represents the vertical component of relative vorticity. $L_d$ is the Rossby deformation radius, a measure of atmospheric stratification. The parameter $\gamma$ denotes the second derivative of the background planetary vorticity with respect to latitude \cite{Nof-1990:Modons}. The term $F\eta$ represents the vorticity injection rate due to stochastic convective motions; $F$ represents the forcing amplitude, and $\eta$ is a normalized stochastic forcing with a primary spatial scale of $L_f$.  The term $\mu q$ represents linear damping of PV, which may arise from turbulent mixing, radiative cooling, or Ohmic dissipation in deeper conductive layers \cite{Thomson-McIntyre-2016:unearthly,Liu-Goldreich-Stevenson-2008:constraints}.

\subsection*{Parameterization of the stratiform instability}

Here, we consider a moist-convective feedback associated with stratiform instability. It has been proposed that, on Jupiter, the low-pressure regions within cyclones may displace isentropic surfaces upward near the water-cloud layer (Fig.~\ref{fig:schematics}). This would locally reduce convective inhibition (CIN) and enhance moist convection \citep{Dowling-Gierasch-1989:moist,Mapes-2000:convective,Juckes-Smith-2000:convective,Thomson-McIntyre-2016:unearthly,Fletcher-etal-2017:moist}. In terrestrial meteorology, this process is known as stratiform instability \citep{Mapes-2000:convective}. 
We note that stratiform instability remains a working hypothesis for Jupiter and has not been confirmed observationally.

We incorporate stratiform instability into \eqref{eq:PVeq} in two steps. First, we estimate how CIN, denoted by $C$ hereafter, responds to cyclonic motion. Second, we consider how the changed CIN affects the strength of convective forcing.

In a conditionally unstable system, the convective inhibition $C$ measures the energy required for an air parcel launched from the base level $H_s$ to reach the level of free convection by overcoming its negative buoyancy. Because latent heat release and removal of high-molecular-weight condensates (ammonia and water vapour) rapidly increase parcel buoyancy upon reaching the lifting condensation level (LCL), we truncate the integration of the negative buoyancy work beyond the LCL, as sketched in Fig.~\ref{fig:schematics}c. In the presence of a vortex, the ambient temperature profile is perturbed through the vortex's warm- or cold-core structure, thereby modifying $C$, as illustrated in Fig.~\ref{fig:schematics}c. To first order, this dependence of $C$ on the vortex streamfunction $\psi$ can be expressed as a linear function:
\begin{equation}
C(\psi) = C_0 + \frac{\partial C}{\partial \psi}\psi,
\label{eq:C-psi}
\end{equation}
where $\partial C/\partial \psi>0$. For cyclones, $\partial_\psi C\cdot\psi<0$ because of the negative temperature perturbation, allowing more air parcels to reach the cloud base. As the system evolves, a background component of $\psi$ develops with the formation of zonal jets, producing a global temperature anomaly. To remove this contribution from \eqref{eq:C-psi}, we apply a high-pass filter to $\psi$ before calculating $C$. The procedure is described in the SI section ``Bridging streamfunction to local temperature anomaly.''

We next consider how CIN translates into the forcing amplitude $F$. The forcing $F$ is assumed to be proportional to the upward mass flux of moist air from the deep interior to the cloud base. With a stably stratified sub-cloud layer, only air parcels with sufficient kinetic energy to overcome the CIN can reach the cloud base and contribute to this upward mass flux. We therefore seek to estimate the fraction of air parcels whose kinetic energy exceeds the CIN. Following \cite{Mapes-2000:convective}, we assume that the turbulent kinetic energy of air parcels in the sub-cloud layer follows a Boltzmann-like exponential distribution, $P(E) = se^{-s E}$, where $\int_0^{\infty} P(E)\,dE=1$, and $s^{-1}$ is the mean turbulent kinetic energy. The fraction of air parcels that overcome the CIN is therefore
\begin{equation}
    F\propto \int_C^\infty P(E)\,dE = e^{-s C}.
    \label{eq:fraction}
\end{equation}

Combining \eqref{eq:C-psi} and \eqref{eq:fraction}, we obtain a relationship between the convective forcing amplitude $F$ and the large-scale streamfunction $\psi$. To facilitate scaling analysis, we normalize $F$ so that the total energy injection rate remains fixed at $\langle\epsilon\rangle$, following Shi \& Kang \cite{Shi-Kang-2026}. This normalization provides the convenience that the actual energy injection rate will not be affected by $\alpha$. It is also physically reasonable: when enhanced convection transmits more heat upward through cyclonic regions, less heat remains to be transmitted through anticyclonic regions, so the overall convective energy input stays roughly the same on average over space and time. In addition, we account for the saturation of convective activity once CIN is fully removed---i.e., after $C$ becomes negative, convection no longer responds to further increases in vortex strength. This yields the final expression for $F$:
\begin{equation}
    F = A_0\sqrt{\langle\epsilon\rangle}
    \frac{e^{\min(-s C_0-s\partial_\psi C\cdot\psi,\,0)}}
    {\sqrt{\langle e^{\min(-2s C_0-2s \partial_\psi C\cdot\psi,\,0)}\rangle}},
    \label{eq:F-psi-cap}
\end{equation}
where $A_0$ is a constant determined by the forcing scale $L_f$ and the forcing renewal timescale \cite{Shi-Kang-2026}.

\subsection*{Parameter space}

Shi and Kang \cite{Shi-Kang-2026} explored the parameter space spanned by the two dimensionless parameters intrinsic to \eqref{eq:PVeq}. Here, accounting for the stratiform instability feedback introduces two additional parameters, $s C_0$ and $s\partial_\psi C$, expanding the parameter space to four dimensions. The four independent dimensionless parameters are:
\begin{subequations}
\label{eq:params}
\begin{align}
    & \frac{L_d}{L_\gamma} \equiv \frac{L_d}{C_\gamma \bar\epsilon^{1/8} \gamma^{-3/8}}, \\
    & \frac{L_\mu}{L_\gamma} \equiv \frac{C_\mu \bar\epsilon^{1/2} \mu^{-3/2}}{C_\gamma \bar\epsilon^{1/8} \gamma^{-3/8}}, \\
    & \alpha \equiv s\partial_\psi C\bar\epsilon^{2/3}\mu^{-2/3}\gamma^{-1/3}, \\
    & B \equiv s C_0.
\end{align}
\end{subequations}
Here, $\alpha$ is nondimensionalized using the length scale $L=\bar\epsilon^{1/6}\mu^{-1/6}\gamma^{-1/3}$ and the time scale $\tau=\bar\epsilon^{-1/3}\mu^{1/3}\gamma^{-1/3}$. The corresponding dimensional parameter, $s\partial_\psi C$, is denoted by $\alpha^*$.

A four-dimensional parameter space is too large to explore thoroughly. As shown in the SI section ``Convective inhibition and its response to a vortex'', both $C_0$ and $\partial_\psi C$ can be estimated from existing observations; consequently, $\alpha$ and $B$ depend on the single unknown parameter $s$. One of these two dimensionless parameters can therefore be treated as dependent, reducing the parameter space to three dimensions. We choose $B$ as the dependent parameter:
\begin{equation}
B = \alpha C_0(\partial_\psi C)^{-1}\bar\epsilon^{-2/3}\mu^{2/3}\gamma^{1/3}.
\end{equation}
Here, $C_0$ and $\partial_\psi C$ depend on the thermal structure of Jupiter's weather layer, which is not well constrained, so the parameter $B$ is uncertain. To assess this uncertainty, we perform sensitivity tests on $B$ in the SI and find that when holding $\alpha$, increasing $B$ produces a similar outcome as increasing $\alpha$. We therefore hold $B$ fixed and focus on the influence of $\alpha$ in the remainder of this work.


We exclude $L_f$ from the main parameter survey and set it to eight times the grid spacing, denoted by $L_{f,\rm ref}$, in most experiments to separate the forcing scale from the dominant dynamical scales. Sensitivity tests presented in the SI show that, in the vortex-soup regime, changing $L_f$ has little effect on cyclone--anticyclone asymmetry, inter-vortex distance, or migration rate. The vortex-crystal regime is more sensitive to $L_f$: among the values tested, the crystal appears most regularly organized at $L_f=2L_{f,\rm ref}$. We therefore use $L_f=2L_{f,\rm ref}$ for experiments in this regime. 

Table~S1 summarizes the parameter combinations for the 56 experiments, in which $L_\mu/L_\gamma$ and $L_d/L_\gamma$ are sampled evenly over $[0.44,1000]$ and $[0.022,4]$, respectively. These ranges cover the regime that may be relevant to Jupiter while keeping the computational cost manageable (experiments become increasingly expensive toward higher $L_\mu/L_\gamma$ and lower $L_d/L_\gamma$). Specifically, the kinetic energy generated by a 7~W/m$^2$ internal heat loss requires $L_\gamma\lesssim 1500$km \cite{ZengEtAl2026}. On the other hand, the deformation radius $L_d$ is estimated to be 200--300km over the polar regions based on vortex drift speed \citep{Gavriel-Kaspi-2025:dynamical}, and $\mathcal{O}(1500)$~km at low latitudes based on various methods \citep{Ingersoll-Kanamori-1995:Waves,Read-Gierasch-Conrath-2006:Mapping,Read-Young-Kennedy-2020:turbulent}. 
For each $(L_\mu/L_\gamma,\ L_d/L_\gamma)$ combination, we gradually increase the $\alpha$ value until the model begins to produce severe grid-scale fluctuations due to numerical instability. 


\subsection*{Model geometry}

We use the PDE solver GeophysicalFlows.jl \cite{Constantinou-Wagner-Siegelman-et-al-2021:geophysicalflows} to solve the governing equation \eqref{eq:PVeq} in a doubly periodic domain centered on the north pole. In all experiments, we set the domain side length to $\min(24L_{\rm Rh},200L_d)$. Here, the Rhines length $L_{\rm Rh}$ characterizes the equilibrium jet spacing and the condensation scale of two-dimensional turbulence under the $\beta$ effect \cite{Rhines-1975:waves}. This choice sets the domain large enough for motions to grow towards the Rhines scale while retaining the resolution needed to study dynamics near $L_d$ when the deformation radius is small. Nevertheless, we test sensitivity to $L_{\rm dom}$ and present the results in SI section ``Sensitivity tests on domain size''. To prevent vortices from re-entering the domain through the periodic boundaries, we impose an artificial damping layer beyond a radius of $0.45L_{\rm dom}$ from the pole, where $L_{\rm dom}$ denotes the side length of the square domain, following the same approach as in \cite{Shi-Kang-2026}.

\begin{figure}[ht!]
\centering
\includegraphics[width=\linewidth]{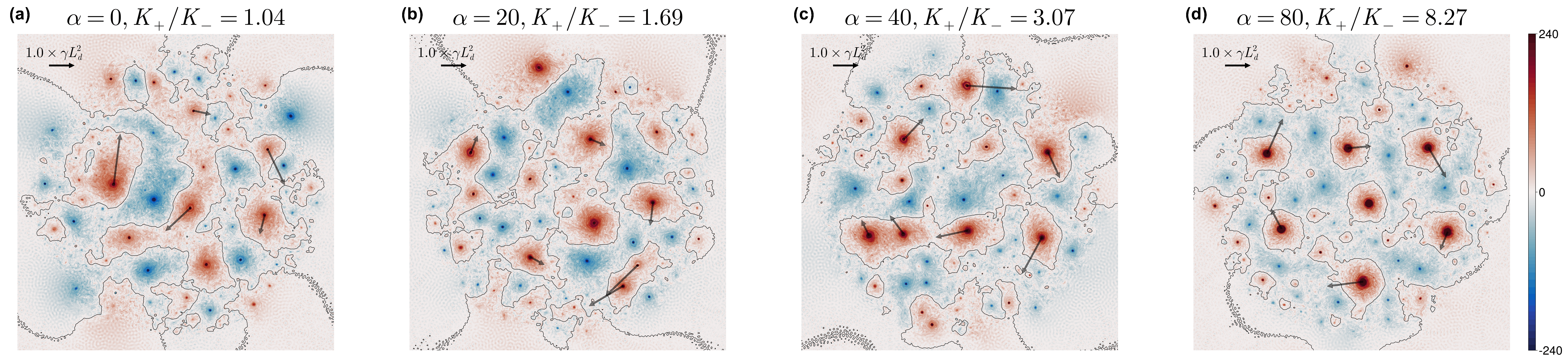}
\caption{Snapshots of potential vorticity ($q$) for different $\alpha$ values, with $L_\mu/L_\gamma=44$, $L_d/L_\gamma=0.068$. Red and blue shading indicates the PV distribution, black solid contours indicate $\psi=0$, and the arrows indicate the migration speeds of each vortex center diagnosed through vortex tracking over a course of 7 years. 
The values of $q$ in this figure are normalized by the time scale $\tau=\bar\epsilon^{-1/3}\mu^{1/3}\gamma^{-1/3}$, as described in Section ``Parameter space''. 
The time evolution of the $q$ field in panel (d) from rest to a stable vortex crystal is available in Movie~S1. The time-evolution of energy and enstrophy for this experiment is shown in Fig.~S13.
}
\label{fig:result-PV-distribution-1}
\end{figure}

\section*{Convective feedback fixes the vortex soup solution by introducing cyclone-anticyclone asymmetry}\label{sec:scaling}




The vortex soup solutions typically emerge in randomly forced QG system when $L_d$ is on the small end ($L_d<L_\gamma$) \cite{Brueshaber-Sayanagi-Dowling-2019:dynamical, Shi-Kang-2026}. The small deformation radius inhibits vortex-vortex interaction and the planetary beta-drift, slowing down the vortex drift speed to approach the observed values \cite{Gavriel-Kaspi-2025:dynamical}. However, unless a vortex crystal structure is prescribed in the initial condition, the flow pattern that emerges from random forcing in a dry model typically features a mixture of cyclones and anticyclones of equal strength. Our goal here is to test whether convective feedback can fix the vortex soup solution by introducing cyclone-ancyclone asymmetry.

As Fig.~\ref{fig:result-PV-distribution-1} shows, increasing $\alpha$, the parameter controlling the strength of the convective feedback, makes cyclones stronger and more coherent, whereas anticyclones become weaker and more diffuse. This produces an increasingly pronounced cyclone--anticyclone asymmetry. Meanwhile, the cyclones' drift speed and the inter-vortex distance remain nearly unchanged. This example demonstrates that introducing the convective feedback has the potential to reproduce a Jupiter-like polar vortex crystal. Over long integrations, vortices occasionally merge and regenerate, but their total number fluctuates only modestly, consistent with observations \cite{Mura-Adriani-Bracco-et-al-2021:oscillations} (Movie~S1). Enlarging the computational domain, however, allows more vortices to form (Movie~S2). These vortices spread over the entire domain, with neither zonal jets nor a central polar vortex present, at odds with observations \cite{Orton-etal-2017:first}. Therefore, reproducing the observed vortex crystal and jet structure may require at least one model parameter to vary with latitude. One possibility is that stratification increases from the pole toward midlatitudes, perhaps associated with the poleward weakening of solar insolation. This would cause the deformation radius to increase rapidly away from the pole, facilitating jet formation there.

To identify the most plausible parameter combinations, we next establish scaling laws for four aspects of the solution: the cyclone--anticyclone asymmetry, the vortex drift speed, the inter-vortex distance, and the root-mean-square wind speed.

\subsection*{Cyclone--anticyclone asymmetry}

\begin{figure}[ht!]
\centering
\includegraphics[width=0.7\linewidth]{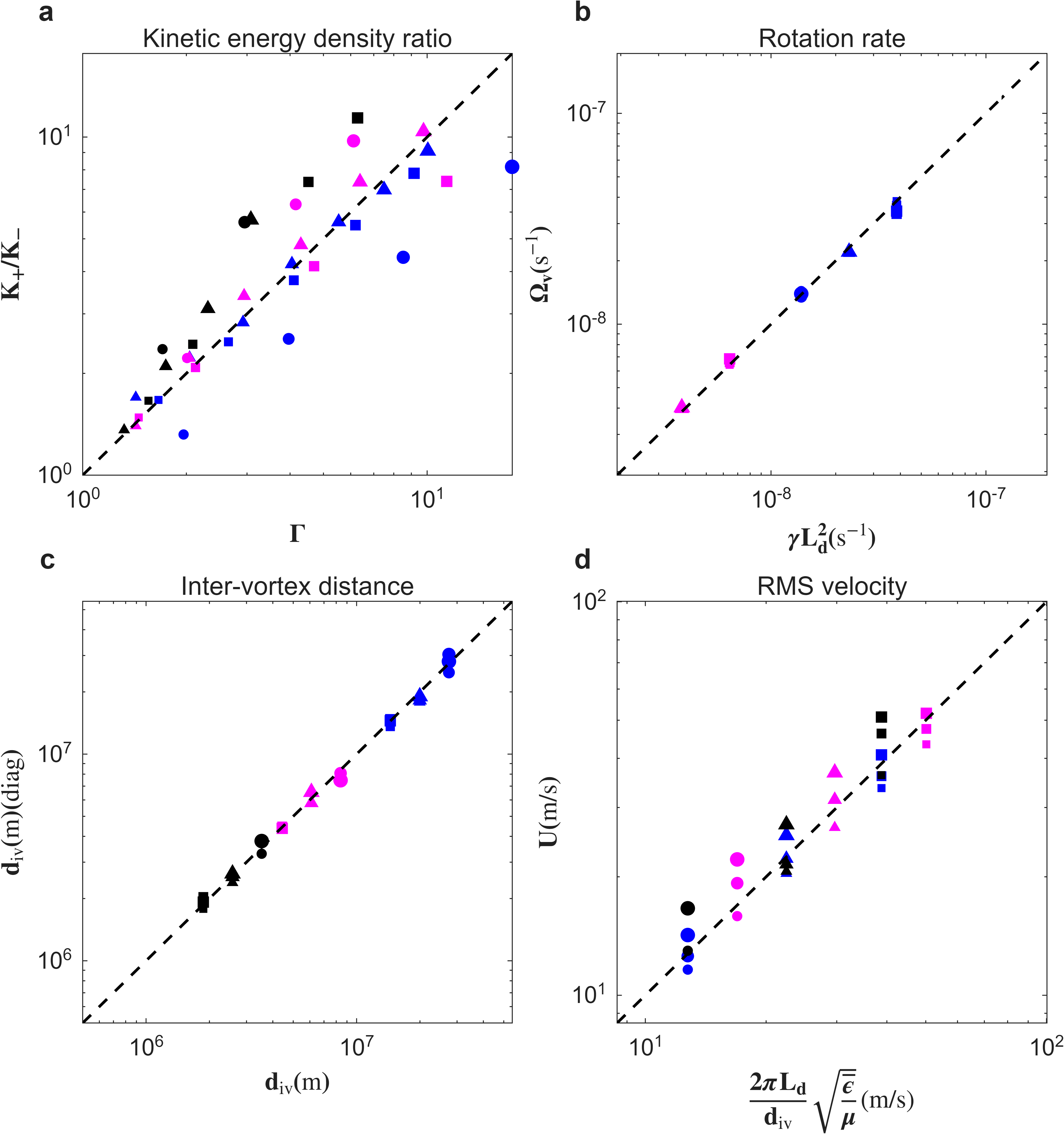}
\caption{A comparison between the diagnosed (vertical axis) and predicted (horizontal axis) values of four observable quantities in our experiments. (a) Cyclone--anticyclone kinetic-energy-density ratio $K_+/K_-$. (b) Rotation rate $\Omega_v$. (c) Inter-vortex distance $d_{\rm iv}$. (d) Root-mean-square velocity $U$. Markers with different shapes, colors, or sizes indicate different experiments, as summarized in Table~S1.
The dimensional values of these quantities are obtained by setting $\bar\epsilon/\mu=10^4~\mathrm{m^2\,s^{-2}}$ and $\gamma=7\times10^{-20}~\mathrm{m^{-2}\,s^{-1}}$ in all experiments to optimize visualization.}
\label{fig:scalings}
\end{figure}

The asymmetry between cyclones and anticyclones can be understood as follows: cyclones reduce CIN locally, inducing stronger convective forcing than in anticyclonic regions. To quantify this asymmetry, we partition the domain into cyclonic and anticyclonic regions along the $\psi=0$ isoline and estimate their kinetic-energy-density ratio. In the experiments shown in Fig.~\ref{fig:result-PV-distribution-1}, the measured kinetic-energy-density ratio increases from 1.04 to 1.69, 3.07, and 8.27 as $\alpha$ increases from 0 to 20, 40, and 80.

To derive a scaling law governing the dependence of this asymmetry on $\alpha$, we consider the energy budgets of the cyclonic and anticyclonic regions separately. Neglecting the weak energy exchange across the cyclone--anticyclone boundary, the equilibrium energy density in each region is proportional to the local energy-injection rate:
\begin{equation}
    \frac{e_+}{e_-} = \frac{e^{\min(-2B+2\alpha L_d\sqrt{e_+},0)}}
    {e^{\min(-2B-2\alpha L_d\sqrt{e_-},0)}}.
    \label{eq:eratio}
\end{equation}
A more detailed calculation is presented in the SI (``Energy-density ratio of cyclones and anticyclones'').

Furthermore, taking advantage of the conservation of the domain-integrated PV leads to an estimate of the cyclone-to-anticyclone area ratio:
\begin{equation}
    \frac{D_+}{D_-} \approx \sqrt{\frac{e_-}{e_+}}.
    \label{eq:D-e}
\end{equation}
As the energy ratio $e_+/e_-$ increases with $\alpha$, the area ratio $D_+/D_-$ indeed decreases as predicted, which is evident in Fig.~\ref{fig:result-PV-distribution-1}.

Since only the kinetic-energy component is observable, we derive a scaling for the kinetic-energy-density ratio $\Gamma$:
\begin{equation}
\Gamma=\frac{\langle|\nabla\psi|^2\rangle_{D_+}}{\langle|\nabla\psi|^2\rangle_{D_-}} \approx \frac{D_-}{D_+}\frac{\langle\psi^2\rangle_{D_+}}{\langle\psi^2\rangle_{D_-}} \approx \frac{D_-}{D_+}\frac{e_+}{e_-}.
    \label{eq:K-ratio}
\end{equation}
Combining Eqs.~(\ref{eq:eratio}), (\ref{eq:D-e}), and (\ref{eq:K-ratio}) fully determines $\Gamma$ from the model parameters. Fig.~\ref{fig:scalings}a compares the predicted value of $\Gamma$ with $K_+/K_-$ diagnosed from all Solution 1 cases and shows good agreement overall. 

\subsection*{Vortex migration}

The vortex drift speed $V_i$ is determined jointly by three contributions: advection by the zonal-mean background flow, advection by the other vortices, and a self-induced tendency, which arises when the vortex's own flow advects the PV of the background and of neighboring vortices \cite{Flierl-1977:Gulf,McWilliams-Flierl-1979:evolution,Sutyrin-1988:Motion,Sutyrin-Flierl-1994:intense,Korotaev-Fedotov-1994:dynamics}:
\begin{multline}
    \partial_t q_i = -V_i \cdot \nabla q_i = -J(\psi_m + \psi_o, q_i) 
    - J(\psi_i, \gamma r^2/2 + q_m + q_o)\\
    \sim -J(\psi_i, \gamma r^2/2) \quad\text{if}~ L_d/L_\gamma \ll 1,
    \label{eq:omega-v}
\end{multline}
where $J(A,B)\equiv (\nabla A\times \nabla B)\cdot \hat{z}$ is the Jacobian operator, and subscripts $(\cdot)_m$, $(\cdot)_o$, and $(\cdot)_i$ denote contributions from the zonal-mean background flow, other vortices, and the vortex of interest, respectively. 

In the vortex soup regime, mutual advection between vortices is largely suppressed because the inter-vortex distance substantially exceeds $L_d$, the threshold for strong vortex--vortex interaction \cite{Polvani-Zabusky-Flierl-1989:two,Shi-Kang-2026}. Although cyclones intensify as $\alpha$ increases, the enhancement is insufficient to produce vortex-vortex interaction; a quantitative assessment is provided in the SI section ``Contribution of advection by neighboring vortices to vortex migration''.
Furthermore, zonal flow does not develop here either, because energy injected at small scales is preferentially used to form vortices before cascading to $L_\gamma$, where zonal jets would otherwise emerge \cite{Shi-Kang-2026}.
The only remaining term in \eqref{eq:omega-v}, $-J(\psi_i,\gamma r^2/2)$, corresponds to advection of planetary PV by the vortex itself. As cyclonic motion advects high-PV fluid from the poleward side toward the western side of the cyclone, the cyclone centre drifts westward---the primary component of $\beta$-drift \cite{Sutyrin-1988:Motion,Sutyrin-Flierl-1994:intense}.

Flierl \cite{Flierl-1977:Gulf} derived an analytical prediction for the $\beta$-drift speed under the assumption that $L_d$ is significantly smaller than the vortex radius, a condition approximately satisfied here. The drift speed $u_\beta$ and its equivalent rotation rate about the pole $\Omega_\gamma$ are
\begin{equation}
    u_\beta \approx \beta L_d^2,\quad \Omega_\gamma \approx \gamma L_d^2.
    \label{eq:beta-drift}
\end{equation}
Neither quantity depends explicitly on $\alpha$, consistent with the insensitivity of vortex migration speed to $\alpha$ seen in Fig.~\ref{fig:result-PV-distribution-1}. 

The drift behavior of polar vortices has previously been investigated systematically by Gavriel and Kaspi \cite{Gavriel-Kaspi-2023:westward,Gavriel-Kaspi-2025:dynamical}. By treating the entire vortex crystal as a single body, they showed that the crystal's long-term migration is controlled by the $\beta$ drift, although vortex--vortex interactions can introduce additional oscillatory motions that deform the crystal arrangement.

Within the vortex soup parameter regime ($L_d<L_\gamma$), we diagnose each vortex's drift speed by tracking its center location over a period of $7$~yrs. 
Figure~\ref{fig:scalings}b compares these diagnosed migration rates with the prediction from \eqref{eq:beta-drift}, showing good agreement. 

\subsection*{Inter-vortex distance}

Following \cite{Shi-Kang-2026}, when $L_d\ll L_\gamma,L_\mu$, the inter-vortex distance $d_{\rm iv}$ scales with the energy-containing scale of two-dimensional turbulence, which can be estimated from the energy budget:
\begin{equation}
    \int_{k_{\rm iv}}^\infty C\epsilon^{2/3}
    (k^{-5/3}+k^{-11/3}L_d^{-2})\,dk = \langle e\rangle=\bar\epsilon/2\mu.
    \label{eq:d_iv}
\end{equation}
The left-hand side estimates the total energy density by integrating the energy spectrum from the largest energy-containing scale (corresponding to $k_{\rm iv}$) to arbitrarily small scales ($k\rightarrow\infty$). This must equal the equilibrium energy level set by the balance between the energy injection rate $\bar\epsilon$ and the damping rate $\mu$ on the right-hand side. The spectral form differs from the standard Kolmogorov $-5/3$ law due to the finite deformation radius $L_d$, as first described by \cite{Larichev-McWilliams-1991:weakly}. Given $\bar\epsilon$, $\mu$, and $L_d$, $k_{\rm iv}$ can be solved implicitly from \eqref{eq:d_iv}.

Since both the spectrum and the equilibrium energy are independent of $\alpha$, the inter-vortex distance remains unchanged as $\alpha$ increases, consistent with Fig.~\ref{fig:result-PV-distribution-1}. Fig.~\ref{fig:scalings}c confirms this, showing good agreement between the predicted $d_{\rm iv}$ and that diagnosed from the simulations.

\subsection*{Average wind speed}

When $L_d\ll L_\gamma,L_\mu$, the root-mean-square wind speed $U$ scales with
\begin{multline}
    U=\sqrt{\langle|\nabla\psi|^2\rangle_{D_0}} \approx \frac{2\pi}{l}\sqrt{\langle\psi^2\rangle} \approx \frac{2\pi}{l}L_d\sqrt{\langle e\rangle} \\ \sim \frac{\pi}{d_{\rm iv}}L_d\sqrt{\frac{\bar\epsilon}{\mu}},
    \label{eq:U2}
\end{multline}
where the characteristic length scale $l$ scales with the inter-vortex distance. Combining \eqref{eq:d_iv} and \eqref{eq:U2} yields the scaling for $U$. Fig.~\ref{fig:scalings}d compares the diagnosed $U$ with this scaling and shows good overall agreement.


\subsection*{Model parameters constrained by Jupiter observations}

Our exploration of the three-dimensional parameter space reveals how polar vortex structure depends on the four model parameters $\bar\epsilon$, $\mu$, $L_d$, and $\alpha^*$. A natural next question is where Jupiter lies in this parameter space and what that location reveals about its internal properties. In this section, we derive observational constraints on the four quantities $\Gamma$, $\Omega_v$, $d_{\rm iv}$, and $U$ from recent Juno measurements, then use these constraints to infer the four model parameters for Jupiter. Measurements of these four quantities and their observational uncertainties are provided in the SI section ``Observational constraints''.

\begin{figure}[ht!]
\centering
\includegraphics[width=0.7\linewidth]{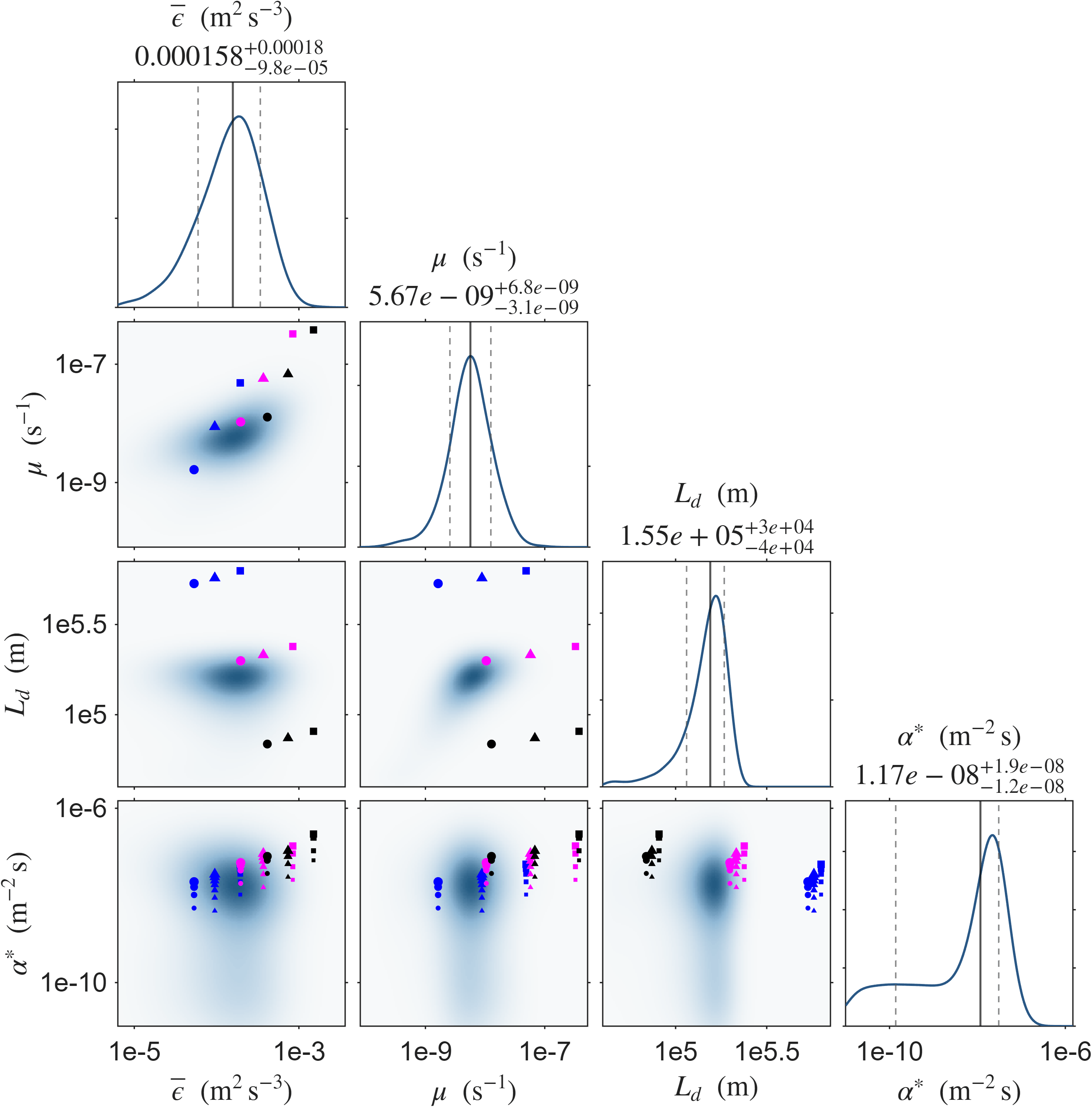}
\caption{The four model parameters, $\bar\epsilon$, $\mu$, $L_d$, and $\alpha^*$, inferred by an MCMC retrieval from the four observable quantities. All axes are displayed in physical units. Diagonal panels show the posterior distributions of the four parameters, with solid vertical lines marking the posterior medians and dashed lines indicating the 16th and 84th percentiles. Markers indicate experiments, and their shape, color, and size all have the same definitions as in Fig.~\ref{fig:scalings} and Table~S1.}
\label{fig:params}
\end{figure}

The scaling laws obtained in this section reveal how the model parameters $\bar\epsilon$, $\mu$, $L_d$, and $\alpha^*$ affect the observables: the vortex characteristic speed $U$, the inter-vortex distance $d_{\rm iv}$, the cyclone--anticyclone asymmetry $\Gamma$, and the vortex drift speed $\Omega_v$. 
Here, we use these relationships to convert observations of Jupiter's north polar flow field into the most plausible model parameters for reproducing such a vortex system within the vortex-soup regime ($L_d<L_\gamma$). A similar analysis can be carried out for the southern hemisphere once wind-field observations become available.

We proceed in two steps. First, we use the reconstructed flow field to constrain $\Gamma$, $\Omega_v$, $d_{\rm iv}$, and $U$. Details of this step and the associated uncertainties are provided in the SI section ``Observational constraints''.
Second, we perform a Markov chain Monte Carlo (MCMC) retrieval to determine Jupiter's $\bar\epsilon$, $\mu$, $L_d$, and $\alpha^*$. A description of the algorithm is presented in the SI (``MCMC retrieval algorithm''). 
Note that, although the scaling laws are validated against only 42 simulations in the vortex-soup regime ($L_d<L_\gamma$), they are more than simple empirical fits. Each relationship follows from a dominant balance among terms in the governing equations that is expected to hold throughout this regime, so the simulations test the predicted parameter dependence instead of defining it. The retrieval can therefore be applied with more confidence than an interpolation among the 42 simulations, including at parameter combinations that were not simulated explicitly, provided they remain in the same regime.

The resulting posterior distributions of the four parameters are summarized in Fig.~\ref{fig:params}, with median values of
\begin{equation}
    \begin{split}
    & \bar\epsilon=1.58\times10^{-4}~\mathrm{m^2\,s^{-3}},\;\mu=5.67\times10^{-9}~\mathrm{s^{-1}}, \\
    & L_d=155~\mathrm{km},\;\alpha^*=1.17\times10^{-8}~\mathrm{m^{-2}\,s},
    \end{split}
    \label{eq:four-params}
\end{equation}
where $\alpha^*=s\partial_\psi C$, which can then be translated into the sub-cloud average turbulent energy $s^{-1}$:
\begin{equation}
    s^{-1} = \partial_\psi C/\alpha^* = 3.0\times10^{4}~\mathrm{m^2\,s^{-2}}.
    \label{eq:turbulence-energy}
\end{equation}

Gavriel and Kaspi \cite{Gavriel-Kaspi-2025:dynamical} inferred a deformation radius of approximately $192\,\mathrm{km}$ for Jupiter's north pole, using an MCMC retrieval based on a large ensemble of dry-QGPV simulations. This is comparable to our estimate of $155\,\mathrm{km}$ in \eqref{eq:four-params}. The small discrepancy may arise because the vortex velocity profiles that emerge in our simulations differ from the observed profile used to prescribe the vortices in \cite{Gavriel-Kaspi-2025:dynamical}. In Fig.~S14, we diagnose the time-averaged vorticity profile of the major vortices from an experiment and compare it with that of \cite{Gavriel-Kaspi-2025:dynamical}. The comparison shows that our vortices are more coherent and exhibit weaker mutual advection, so to achieve the same migration speed, our estimate of $L_d$ is slightly smaller. The set of experiments with the most plausible $L_{\mu} / L_{\gamma}, L_d / L_{\gamma}$ is presented in Fig.~\ref{fig:result-PV-distribution-1} as our example case.


\begin{figure}[ht!]
\centering
\includegraphics[width=\linewidth]{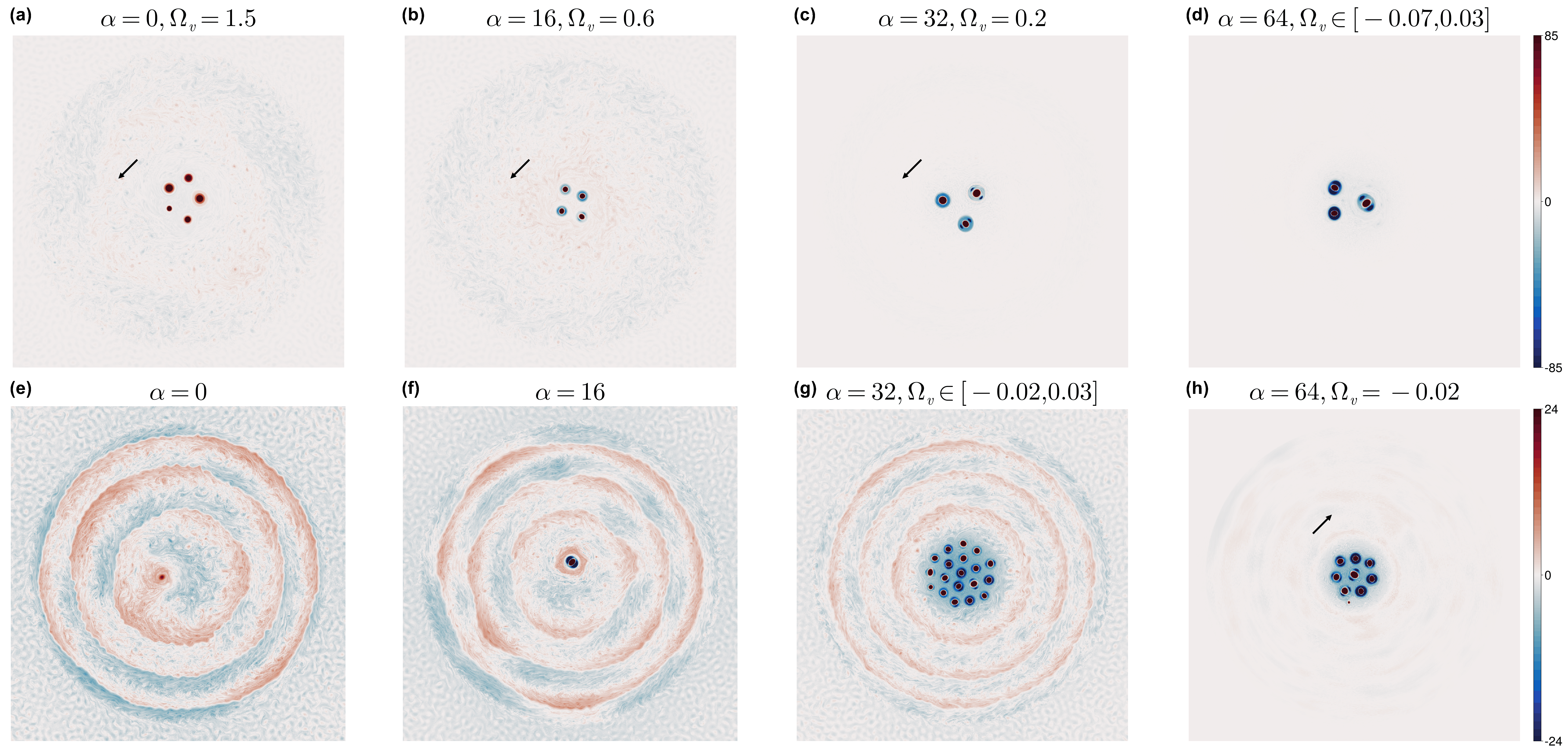}
\caption{Snapshots of potential vorticity ($q$) at different values of $\alpha$ in two experiments in the intermediate- and large-$L_d$ vortex crystal regime. The top row has $L_\mu/L_\gamma=200$ and $L_d/L_\gamma=4$, the bottom row has $L_\mu/L_\gamma=100$ and $L_d/L_\gamma=2$. In the top row, the domain size is set to 12$L_{\rm Rh}$ (half the default value) due to the high computational cost of evolving the strong initial turbulence. The shading and $q$ values follow the conventions of Fig.~\ref{fig:result-PV-distribution-1}. Black arrows indicate the direction of vortex crystal rotation, with its amplitude given in each panel title. Arrows are omitted where the crystal changes its direction of rotation over time. Movies~S3, S4, and S5 show the full evolution of the $q$ field in panels (g), (h), and (d), respectively; Figs.~S10, S11, and S12 show the corresponding evolution of energy and enstrophy.
}
\label{fig:result-PV-distribution-2}
\end{figure}

\section*{Convective feedback slows down the vortex migration of the vortex crystal solution}

Vortex crystal solutions typically emerge in QG systems when $L_d$ is large and the initial energy, seeded as random noise with a spectrum concentrated near $k_f$, is sufficiently high \cite{Siegelman-Young-Ingersoll-2022:polar, Chen-Ingersoll-Li-2024:jupiter, Shi-Kang-2026}. In such a system, the initial energy cascades to large scales and forms zones of uniform potential vorticity. These zones cut off the $\beta$ effect that drives the poleward drift and merging of cyclones. One issue with this solution is that the surrounding vortices tend to drift quickly in the cyclonic direction, because they are advected by the other cyclones. In observations, however, Jupiter's polar vortices drift slowly in the anticyclonic direction \cite{Mura-Adriani-Bracco-et-al-2021:oscillations}. 

As the top row of Fig.~\ref{fig:result-PV-distribution-2} shows, strengthening the convective feedback slows and even reverses the vortex drift. The underlying mechanism is as follows. The feedback concentrates the random forcing near the cyclones, whereas damping acts uniformly over the domain. As a result, the forced perturbations, whether cyclonic or anticyclonic, cannot travel far from the vortex center before being damped. Because each vortex sustains a local PV gradient, cyclonic perturbations are absorbed into the parent cyclone, whereas anticyclonic perturbations are expelled to form a skirt around it. This anticyclonic shielding layer substantially weakens vortex--vortex advection \cite{Li-Ingersoll-Klipfel-et-al-2020:modeling} and, if sufficiently strong, can reverse the drift to the anticyclonic direction.
In the absence of the convective feedback, anticyclonic perturbations generated far from the cyclones never approach them. Those generated near a cyclone are expelled far enough to merge into the anticyclonic background (Fig.~\ref{fig:result-PV-distribution-2}a), so no shielding layer forms around the cyclone.

The vortex crystal in Fig.~\ref{fig:result-PV-distribution-2}a is, however, still sensitive to the initial condition like those in the dry model \cite{Shi-Kang-2026}. Without sufficiently high initial energy, the system ends up with a single vortex or a vortex pair, as shown in Fig.~S5 in the SI. 

When the deformation radius $L_d$ is comparable to the zonostrophic scale $L_\gamma$, \cite{Shi-Kang-2026} found a transitional regime in which only a single vortex can be sustained regardless of the initial condition, because both Mechanism~1 and Mechanism~2 fail. In this regime, we find that the convective feedback can sustain a vortex crystal by wrapping each cyclone in an anticyclonic shielding layer (see the bottom row in Fig.~\ref{fig:result-PV-distribution-2}). This anticyclonic shielding sustaining the crystal is known to repel neighboring cyclones and prevent them from merging \cite{Li-Ingersoll-Klipfel-et-al-2020:modeling,Gavriel-Kaspi-2021:number}. Unlike the crystal that forms at large $L_d$, the crystal forms at intermediate $L_d$ requires the initial energy to be sufficiently lower than $\bar\epsilon/2\mu$ (Fig.~S5 in the SI). 

To conclude, with the convective feedback, both the observed vortex pattern and drift speed can in principle be reproduced simultaneously. Identifying the parameter combinations that match the observations is nonetheless difficult in this case, because the number of vortices sustained and the strength of the shielding layer are history-dependent. Other than the sensitivity to initial conditions, changing the second convective feedback parameter $B$ or the forcing scale $L_f$ turns out to affect the vortex structures and the vortex drift speeds as well (see Fig.~S2 and Fig.~S3 in the SI). Although we cannot yet pinpoint the exact parameters that produce Jupiter-like vortices, the strong sensitivity of these key observables to the feedback gives us confidence that such a solution exists somewhere in the parameter space.

\section*{Conclusion and discussion}\label{sec:conclusion}

In this work, we explored whether a moist-convective feedback can resolve a key challenge that dry models face in reproducing Jupiter's polar vortices. Matching the observed westward drift of the polar vortices requires a deformation radius $L_d$ too small for dry QG or shallow-water models to sustain a stable, cyclone-dominated vortex crystal \cite{Gavriel-Kaspi-2025:dynamical,Chen-Ingersoll-Li-2024:jupiter,Shi-Kang-2026}. Conversely, reproducing the crystal structure requires an $L_d$ too large for dry models to yield a drift that is slow enough and in the right direction \cite{Gavriel-Kaspi-2025:dynamical}. The convective feedback we considered is based on stratiform instability. In regions dominated by cyclonic motion, the associated thermal anomalies reduce CIN, so a larger fraction of sub-cloud-layer perturbations can develop into convection. This enhances the convective forcing there. We find that the feedback narrows the gap between dry models and observations across the range of $L_d$ we examined:
\begin{itemize}
\item \textbf{Small $L_d$ ($L_d/L_\gamma<1$):} The vortex field needs better organization and cyclone dominance. The feedback concentrates the convective forcing in cyclonic regions, which strengthens cyclones relative to anticyclones and produces a cyclone-dominated crystal structure.
\item \textbf{Large $L_d$ ($L_d/L_{\rm Rh}>1$):} Dry models drift too fast and in the wrong direction. Under the feedback, an anticyclonic shielding layer forms around each cyclone. This layer cuts off the mutual advection between vortices and slows, or even reverses, the drift to be consistent with the observation \cite{Mura-Adriani-Bracco-et-al-2021:oscillations}.
\item \textbf{Intermediate $L_d$ ($L_{\gamma}<L_d<L_{\rm Rh}$):} Dry models sustain only a single vortex, because both multi-vortex mechanisms fail. The feedback restores a multi-vortex crystal by wrapping each vortex in an anticyclonic shielding layer.
\end{itemize}

We then develop quantitative scaling laws for the small-$L_d$ regime. These laws predict four observable quantities from four model parameters: the partition of kinetic energy between cyclonic and anticyclonic regions $K_+/K_-$, the vortex drift rate $\Omega_v$, the inter-vortex distance $d_{\rm iv}$, and the root-mean-square wind speed $U$. We then invert these relationships to constrain the four model parameters for Jupiter's north polar vortices, using Juno measurements. The most likely parameter combination is $\bar\epsilon=1.58\times10^{-4}~\mathrm{m^2\,s^{-3}},\;\mu=5.67\times10^{-9}~\mathrm{s^{-1}}, L_d=155~\mathrm{km},\;\alpha^*=1.17\times10^{-8}~\mathrm{m^{-2}\,s}$. 
For intermediate and large $L_d$, the final vortex configuration and drift behavior are sensitive to various parameters not included in our parameter space, such as the second moist feedback parameter $B$ and the forcing scale $L_f$, making it difficult to obtain quantitative scaling laws.

A Rossby deformation radius of $1000\mbox{-}2000\,{\rm km}$ is estimated by previous studies from the Shoemaker--Levy 9 impact \cite{Ingersoll-Kanamori-1995:Waves}, various instabilities \cite{Read-Gierasch-Conrath-2006:Mapping,Thomson-McIntyre-2016:unearthly}, and the energy transfer spectrum \cite{Read-Young-Kennedy-2020:turbulent}. This range lies in the intermediate $L_d$ regime of this study, where we do find several cyclone-dominated, westward-drifting vortex crystal solutions that resemble Jupiter's, as shown in the bottom row in Fig.~\ref{fig:result-PV-distribution-2}. The intermediate $L_d$ regime is therefore a promising candidate for where Jupiter is located.

We note that the QG framework adopted here is highly idealized and omits several potentially important processes. First, stratiform instability has not been observationally established in Jupiter's polar atmosphere. We use it as a physically motivated representation of a possible feedback between cyclonic circulation and convective forcing, and its strength depends on the vertical temperature structure, parcel source level, and cloud-base dynamics, all of which need stronger observational constraints. Second, the model assumes geostrophic balance and neglects centrifugal effects, which grow with the Rossby number. Because the centrifugal force reinforces the radial pressure gradient in cyclones but opposes it in anticyclones \cite{Scott-2007:coherent}, it may introduce additional cyclone--anticyclone asymmetry. Third, all vortices in our simulations form locally within the polar region, whereas some polar vortices may originate at lower latitudes and migrate poleward \cite{Li-Ingersoll-Klipfel-et-al-2020:modeling}. This is possible in reality but not represented in our modeling framework.

Despite these simplifications, our results show that a simple, physically motivated convective feedback can reconcile the vortex pattern and drift speed that dry models could not match together. They also give a quantitative framework for linking Jupiter's polar vortices to the properties of its deep atmosphere. As Juno continues to observe the poles and constraints on the water-cloud layer improve, this framework can be tested and refined. We hope it will help bring the dynamics of Jupiter's polar vortices into sharper focus.

\subsection*{Supporting Information}

A supporting information (SI) pdf together with 5 movies are available as ancillary files.

\subsection*{Data, Material, and Software Availability}

The Julia package GeophysicalFlows.jl is available at \url{https://github.com/FourierFlows/GeophysicalFlows.jl}. All modified code for GeophysicalFlows.jl and code for the MCMC retrieval algorithm can be accessed through Zenodo at \url{https://doi.org/10.5281/zenodo.21370681}.

\section*{Acknowledgments}

We would like to acknowledge high-performance computing support from the Derecho system (doi:10.5065/qx9a-pg09) provided by the NSF National Center for Atmospheric Research (NCAR), sponsored by the National Science Foundation. We also thank MathWorks for developing MATLAB and its data analysis tools, with which we developed the Markov chain Monte Carlo (MCMC) algorithm used in this study. We acknowledge support from NASA and MIT. W.K. discloses funding from NASA under grant 80NSSC26K0263 and discretionary funds from MIT.

\bibliographystyle{arxiv-etal}
\bibliography{references}

\end{document}